# Evaluation of Virtual Acoustic Environments with Different Acoustic Level of Detail

Stefan Fichna
*Medizinische Physik
and Cluster of Excellence Hearing4all*
Universität Oldenburg
Oldenburg, Germany
stefan.fichna@uol.de

Steven van de Par
*Akustik
and Cluster of Excellence Hearing4all*
Universität Oldenburg
Oldenburg, Germany
steven.van.de.par@uol.de

Stephan D. Ewert
*Medizinische Physik
and Cluster of Excellence Hearing4all*
Universität Oldenburg
Oldenburg, Germany
stephan.ewert@uol.de

*Abstract*— Virtual acoustic environments enable the creation and simulation of realistic and ecologically valid daily-life situations with applications in hearing research and audiology. Hereby, reverberant indoor environments play an important role. For real-time applications, simplifications in the room acoustics simulation are required, however, it remains unclear what acoustic level of detail (ALOD) is necessary to capture all perceptually relevant effects. This study investigates the effect of varying ALOD in the simulation of three different real environments, a living room with a coupled kitchen, a pub, and an underground station. ALOD was varied by generating different numbers of image sources for early reflections, or by excluding geometrical room details specific for each environment. The simulations were perceptually evaluated using headphones in comparison to binaural room impulse responses measured with a dummy head in the corresponding real environments, and partly using loudspeakers. The study assessed the perceived overall difference for a pulse, and a speech token. Furthermore, plausibility and externalization were evaluated. The results show that a strong reduction in ALOD is possible while obtaining similar plausibility and externalization as with the dummy head recordings. The number and accuracy of early reflections appear less relevant, provided diffuse late reverberation is appropriately accounted for.

## I. INTRODUCTION

In psychoacoustics, listening experiments are typically conducted in highly controlled laboratory conditions offering the ability to manipulate and regulate all relevant variables with precision. However, results obtained from such laboratory conditions, which often entail simplified acoustic settings and synthetic stimuli, may not accurately reflect real-life situations. This disparity has become increasingly apparent in studies focused on hearing aid performance and speech intelligibility [1].

To bridge the gap between simplified laboratory measurements and real-life listening and communication situations, complex acoustic environments (CAEs; [2, 3]) can be used in conjunction with appropriate techniques for acoustical reproduction and rendering [4, 5, 6]. Hereby, real-life CAEs [7, 8, 9] provide "ground truth" data, establishing a benchmark against which laboratory-based measurements can be evaluated. One critical acoustical factor in indoor CAEs reflecting typical communication-related situations is the presence of sound reflections and reverberation.

CAEs can be generated in a highly controllable way using room acoustics simulation and virtual acoustics. Several room acoustics simulation approaches exist [e.g., 4, 5, 6, 10, 11] all involving simplifications of the underlying acoustical processes. A high degree of perceptual plausibility can be reached with state-of-the-art approaches [12]. Nevertheless, it remains unclear to which extent the acoustic level of detail (ALOD) within these simulations can be reduced while still being perceptually plausible or achieving a certain agreement of spatial audio quality and speech intelligibility between simulation and the respective real-life CAE. Particularly with regard to interactive real-time applications, a considerable reduction of ALOD in the simulation is desirable.

Regarding acoustic parameters such as reverberation time and speech transmission index (STI, [13]), Abd Jalil et al. [14] have shown that the number of surfaces in the acoustic models could be reduced by as much as 80 %. Furthermore, it has been demonstrated that even with simplification to shoebox geometry [6, 11], perceptual plausibility and a close agreement with measured reference conditions can be achieved [e.g., 12, 15, 16]. In addition to enabling low-latency, interactive CAEs for hearing research, such a highly simplified acoustic simulation is in principle suited to run on mobile hardware, offering possibilities for realistic spatial rendering in hearables and hearing supportive systems [e.g., 17, 18]. Considering a shoebox approximation of the main enclosure, it remains unclear which impact the more detailed simulation of nearby reflecting objects or coupled volumes have on perception.

The goal of this study is to investigate how different ALODs within the room acoustics simulation influence plausibility, the overall perceived difference to a reference auralization, and externalization across three distinct and acoustically diverse rooms, extending a preliminary study with a focus on speech intelligibility and spatial audio quality in [19]. Binaural room impulse responses (BRIRs), captured through recordings in a living room coupled with a kitchen, a pub, and an underground station [7], were used as a reference for headphone auralization. The perceptual assessments of overall perceived difference and externalization were additionally performed in a 3-dimensional loudspeaker array. The Room Acoustics SimulatoR RAZR [6, 11] was used to generate synthetic BRIRs and loudspeaker renderings.

The ALOD in the simulation was systematically changed, ranging from a simple shoebox image source model [20], to using all features of RAZR including simplified effects of scattering and diffraction. Additionally, the effect of simulating nearby finite reflecting surfaces and the simulation of coupled volumes was assessed, depending on the presence of that specific feature in the environment.

This work was funded by the Deutsche Forschungsgemeinschaft, DFG SPP Auditive – Project-ID 444827755 and DFG – Project-ID 352015383 – SFB 1330 C5.

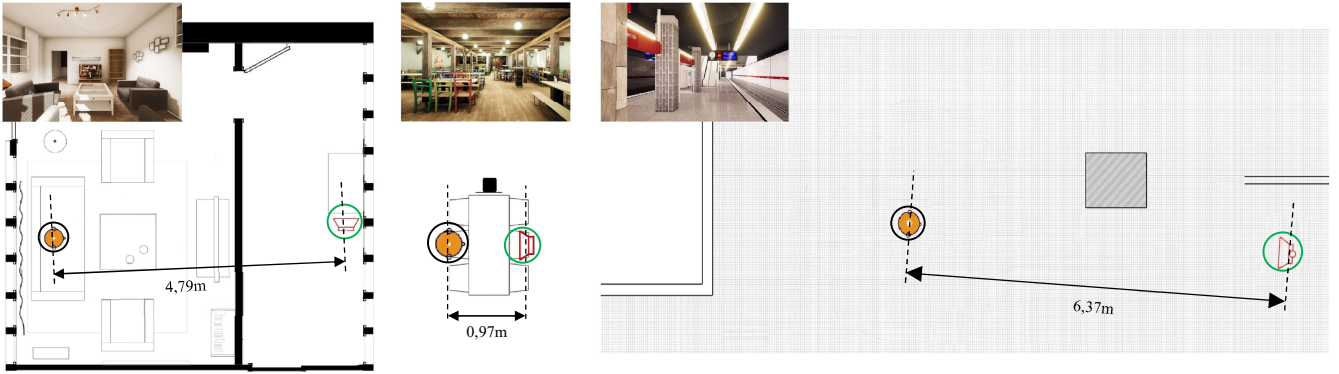

Fig. 1. Layouts of the three scenes in the three different real-world environments: Living room, pub, and underground station (left to right). The upper insets show visual renderings generated by Unreal Engine. The receiver and target positions are marked with black and green circles, respectively. For a scale comparison, the distance between the receiver and the target position is indicated. For the living room, the direct line of sight to the target was obstructed. For the pub and underground (middle and right) only sections of the floorplan are provided, for a full description see [7].

## II. METHODS

### A. Participants

The experiment involved eight participants with normal hearing, aged between 21 and 33 years. All participants had prior experience in psychoacoustic tests and received an hourly compensation.

### B. Acoustic Scenes

Everyday-life acoustic scenes were created based on the dummy head recordings in three real-world acoustic environments depicted in Fig. 1 and described in more detail in [7]. The first scene was set in a furnished living room with dimensions of 4.97 m x 3.78 m x 2.71 m, resulting in a volume of 50.91 m³ and a reverberation time (T30) of 0.54 s. The living room was connected to a kitchen (4.97 m x 2.00 m x 2.71 m) through an open door, with a volume of 26.94 m³ and a reverberation time (T30) of 0.66 s. In this scene, a receiver sat on a sofa in the living room, while the target was in the kitchen and was oriented towards the open door between the two rooms. The direct line of sight to the source was obstructed (see left-hand side of Fig. 1). The path length (through the door) between receiver and target was 5.7 m.

The second scene was located in a pub, which consisted of a single large room with dimensions of 17.76 m x 10.2 m x 2.9 m, resulting in a volume of ~442 m³ and a reverberation time (T30) of 0.7 s. The receiver was located at a table in the pub, with the target on the opposite side of the table, 0.97 m away (see center part of Fig. 1). The pub scene was characterized by a relatively low reverberation time considering its large volume, and several objects in the immediate vicinity of the receiver position, such as the table and a chalkboard to the left at ear level, resulting in pronounced early reflections.

The third scene was set in an underground station, consisting of a large, elongated space (containing the platform and two tracks) of about 120 m x 15.7 m x 4.16 m, with a total volume of about 11000 m³ and a reverberation time (T30) of 1.6 s. The connected underground tunnels and escalators provided additional (coupled) volumes resulting in a dual-slope decay of the late reverberation. The receiver was standing in the middle of the platform and the target signal was located at 6.37 m in front of the receiver (see right-hand side of Fig. 1).

### C. Room acoustic simulation and stimuli

In order to simulate acoustics at varying ALOD, synthetic BRIRs and loudspeaker renderings were generated using RAZR [6, 11]. RAZR employs a "proxy shoebox room" image source model (ISM) for early reflections and a feedback delay network (FDN) for diffuse reverberation using physically-based delay times. RAZR has been assessed in comparison to real acoustic scenes [12, 15, 16] reaching a high degree of agreement and perceptual plausibility.

Here, each scene was simulated with five sets of different features in the room acoustics simulation to vary the ALOD: 1) The highest ALOD used RAZR with all features, including coupled rooms and a third-order ISM for early reflections. The ISM comprises jittering of image source positions to avoid an unrealistic, completely regular reflection pattern, and temporal smearing of early diffuse reflections, simulating scattering and multiple reflections caused by geometric disturbances and objects in the room. In the following, this set will be referred to as RAZR. 2) The level of detail in the early reflections was lowered by reducing the ISM order from three to one, and will be referred to as RAZR-1st-Order. 3) In the third set, referred to as RAZR-Simple, a specific feature of the acoustic simulation was disabled for each room: For the living room, the coupled room simulation was simplified by using two separate simulations for each room. First, only the kitchen was simulated in RAZR and an omnidirectional receiver was placed at the position of the (closed) door. To generate the final simulation, only the living room was rendered with an omnidirectional virtual source placed in the center of the (closed) door, radiating the response of the kitchen. In the pub, nearby reflecting finite surfaces representing the tabletop and chalkboard were removed. In the underground station, a coupled volume representing the underground tunnels and escalators, resulting in a dual-slope characteristic of the energy decay was disregarded in the simulation. 4) The next set with reduced ALOD, referred to as ISM, used a straight-forward implementation of a 15th-order shoebox ISM, thus disregarding effects of sound scattering. This resulted in an unnatural, completely regular pattern of specular reflections. 5) For the externalization experiment, a diotic condition was added, referred to as Diotic. For Diotic headphone tests, the left channel of the measured signal was presented on both ears. For Diotic loudspeaker tests, RAZR was presented by a single loudspeaker in front of the participant.

For the headphone experiments, the above simulated BRIRs and recorded BRIRs from the real rooms [7] (referred to as Measured in the following) were convolved with one of two target signals: Speech material from the matrix speech test Oldenburger Satz Test (OLSA, [21]), which consists of a large number of grammatically correct but semantically unpredictable test sentences constructed from a total of 50 words with ten alternatives for each word type (name-verb-numeral-adjective-object) was used. The sentences were spoken by a male speaker. For the plausibility test eight random sentences were selected. For the evaluation of the overall difference and externalization only a single sentence was used for all presentations to allow for a direct comparison between the BRIR sets with the exact same target signal.

The second target stimulus was a pulse with a pink-colored spectrum. For the plausibility test, the spectrum of the pink pulse was altered within 10 octave bands between 31 Hz and 16 kHz. 8 variations were created by randomly increasing the level of five bands by 6 dB and decreasing the level of the other five bands by 6 dB. This procedure resulted in differently colored pulses with a clearly distinguishable sound impression ranging from "knocking on wood" to "bursting balloon". For overall difference and externalization, the original pulse without spectral modification was used.

*D. Apparatus and procedure*

For the headphone experiments, the listeners were seated in a sound-proof booth with double walls and equipped with Sennheiser HD 650 headphones connected to an RME Fireface UCX at a sampling rate of 44.1 kHz. All listening experiments were performed in Matlab. A computer monitor was placed in front of the participants, and they used a computer mouse and keyboard to provide responses. For loudspeaker measurements (in the overall difference and externalization task) the experiments were performed in the VR lab at the University of Oldenburg which is an anechoic chamber including a 3-dimensional array of 86 Genelec 8030 loudspeakers [see, e.g., 22]. The participants were seated in the centre of the loudspeaker array. The main horizontal ring of the array (radius 2.4 m) consists of 48 loudspeakers and is mounted at about head height of the seated participant. Two horizontal rings with twelve loudspeakers are positioned at an azimuth of - 30° and 30° and two horizontal rings with 6 loudspeakers are positions at an azimuth of - 60° and 60°. Two more loudspeakers are positioned at - 90° and 90° azimuth.

In the plausibility experiment a single, either real (based on recorded BRIRs) or simulated stimulus was presented over headphones. Afterwards the participant was asked "Was the stimulus real or simulated?", resulting in a two-alternative forced choice test. The plausibility test consisted of 6 measurements, one for each scene (living room; pub; underground station) and for the three types of impulse responses used in this test (Measured; RAZR; ISM). For each measurement 16 target signals, 8 sentences and 8 pulses as described in section II.C, were randomly presented three times, resulting in 48 presentations per measurement. Each measurement was performed twice as test and retest. Prior to the measurement, a short training was conducted where the participants listened to some examples: The examples used BRIRs of different positions (than later used during the test) within the same environments. A pulse and a sentence of the speech material were presented. In the training, the dry stimuli were first presented without applying any BRIR. Afterwards, the same stimuli convolved with the Measured and ISM BRIR were presented. During the training, the participants were told which signals were measured or simulated.

To obtain overall difference ratings, a procedure similar to the multi-stimulus test with hidden reference and anchor (MUSHRA; [23]) was applied using headphone and loudspeaker presentation. Listeners rated various stimuli relative to a reference using sliders on the computer screen and were able to listen to the stimuli repeatedly and sort the stimuli according to their ratings. The slider positions reflected a score between 0 and 100, where 100 means "very different to the reference" and 0 means "no difference to the reference". The test signals were convolved with different BRIRs: Measured (only available for headphone presentation), RAZR, ISM, RAZR-Simple, RAZR-1st-Order. RAZR served as the (hidden) reference, whereas ISM effectively took the role of the anchor. A single sentence of the OLSA test and a deterministic pink pulse (with no spectral variation applied) were used as target. This experiment was separated into six different tests consisting of the two different target stimuli (speech and pulse) and three different environments (living room, pub, and underground station). Again, this experiment was measured twice for each participant in a test and a retest.

The last experiment evaluated the perceived externalization. Similar to the procedure in the overall difference measurement a multi-stimulus test was used, however, without any reference. On a scale from 0 to 100 listeners rated the perceived externalization with four verbal descriptors. 100 was labeled "Clearly outside the head", 66 represented "Close to the head", 33 represented "Between the ears" and 0 "Central in the head". The participants could listen to each signal repeatedly. For this experiment, Measured, RAZR, and ISM were used. Similar to the test for the overall difference, this experiment was separated into six different tests consisting of two different type of target signal (speech and pink pulse) and three different environments (living room, pub, and underground station). Again, a test and a retest were conducted. Before the measurement with headphones, a short training was performed. For this, the same environments were used as for the measurement, but with different positions for source and receiver. Stimuli generated with the measured BRIR were presented either diotically or binaurally, to demonstrate the difference between internal and external sounding signals.

For each of the three experiments, the consistency of the responses was examined. The participant's responses for the test and for the retest were separately averaged and the test-retest correlation was calculated. A minimum correlation coefficient of 0.7 was required and participants who did not reach this value were measured a second time, where they all reached the minimum value.

III. RESULTS

*A. Plausibility*

The results of the plausibility test are shown in Fig. 2. The average test-retest correlation across all seven participants was 0.88. The top panel shows results for speech and the bottom panel for the pulse. The color of the violin plots indicates the underlying BRIRs: Here and in the other figures, green shows results for Measured, red shows RAZR, and ISM is shown in blue (from left to right). The open symbols and bars indicate the median and the 25 % and 75% quartiles.

For all conditions plausibility was rated similarly for Measured and RAZR. The highest ratings reached close to 100% plausibility for the pulse in the underground and living room as well as for speech in the pub. The mean difference of the median between Measured and RAZR across all conditions was 3.12 %. ISM was rated considerably less plausible than Measured and RAZR, particularly for the pulse (bottom panel) where the median reached 0 % for the living room and underground. Results with the ISM (blue) show a considerably larger spread for speech than for pulse. Furthermore, for Measured and RAZR, the variability in the data for the underground was considerably larger for speech than for the pulse given that there were listeners who rated neither as plausible. A possible explanation is that rooms with such a high reverberation time as the underground were unfamiliar to those listeners and therefore appeared unnatural. This could also be the reason why for speech in the underground even the ISM was considered plausible multiple times (median = 66.7 %), similar to Measured (median = 70.8 %) and RAZR (median = 70.8 %).

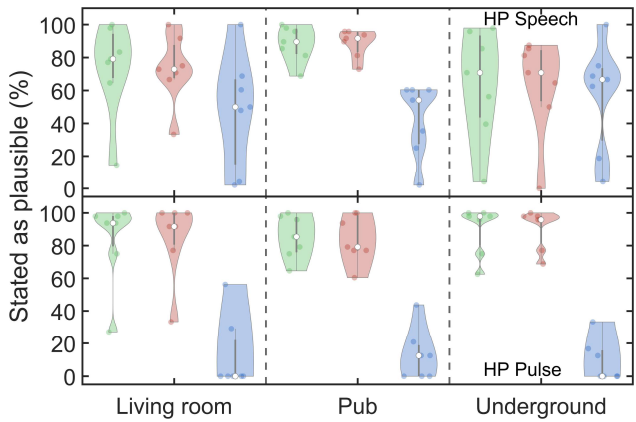

Fig. 2. Percentage of headphone auralizations rated plausible for speech (top panel) and pulse (bottom panel). Green: Measured; red: RAZR; blue: ISM.

Taken together, across all conditions Measured and RAZR were rated plausible for both speech and pulse. ISM was generally rated less plausible, particularly for pulse.

### B. Overall difference

Fig. 3 shows the rated overall difference with RAZR serving as reference for headphone presentation in the upper two rows and for loudspeaker rendering in the lower two rows. Speech and pulse as target stimulus are shown in every other row. The average test-retest correlation was 0.92 and the results were pooled. The colors of the violin plots are the same as for plausibility with two new ALOD conditions added here (from left to right), green: Measured (headphone presentation only), red: RAZR (reference), blue: ISM, black: RAZR-Simple; teal: RAZR-1st-Order. A value of 0 on the y-axis represents no perceptible difference from the reference signal and a value of 100 represents the rating "very large difference from the reference".

The hidden reference (RAZR, red) was mostly detected and rated with no difference, indicating reliability of the results. On average, ISM was rated to have the highest difference to the reference (RAZR). For the pulse, differences were overall rated higher than for speech. This is consistent with the results of the plausibility test, where ISM was more often rated as plausible (similar to Measured) for speech than for pulse. RAZR-1st-Order and RAZR-Simple were also rated very similar to the reference for the underground station. The overall difference for RAZR-Simple was higher for the pub and highest for the living room. The difference between headphone and speaker representation was generally small.

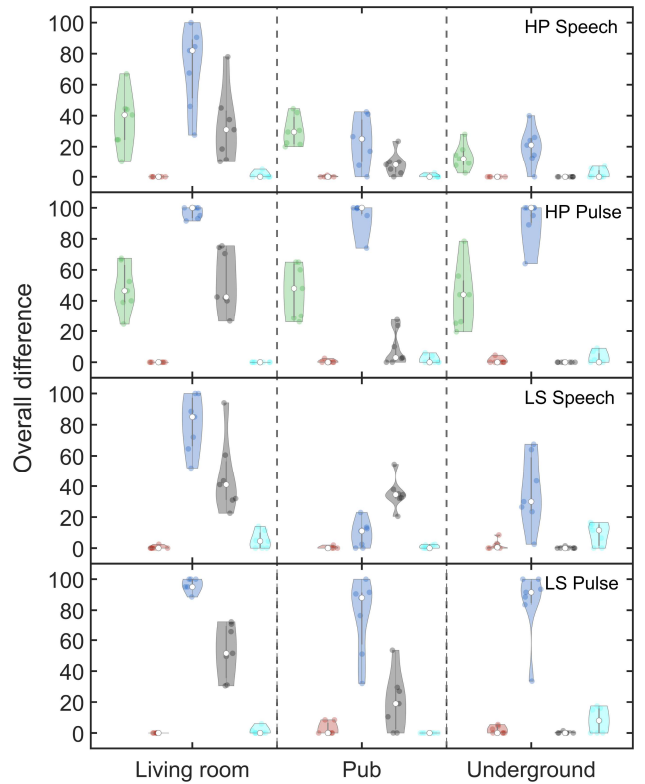

Fig. 3. Results for the overall difference rating for headphones (upper two panels; HP) and loudspeaker (lower two panels; LS). Every other row is for the speech and pulse target. Green: Measured; red: RAZR; blue: ISM; black: RAZR-Simple; teal: RAZR-1st-Order.

One exception was RAZR-Simple for the pub. An explanation might be that the sound reflection from the table which was removed in RAZR-Simple was spatially better resolved in elevation in the loudspeaker array than for headphones. The loudspeakers at $-30°$ elevation might have provided additional spatial cues. In an informal interview some participants stated that they perceived differences between RAZR-Simple and RAZR particular from loudspeakers at $-30°$ elevation. Additionally, the chalk board on the left side of the participant could have added additional spatial cues.

The small difference rating for RAZR-Simple in the underground is likely related to the low level where the dual slope decay occurs at about -15 dB below the peak of the EDC (see Fig. 6 in [7]).

Taken together, differences between Measured and RAZR occurred for the headphone comparison. Overall RAZR-1st-Order was rated quite similar to RAZR.

### C. Externalization

In Fig. 4 the results of the externalization measurement are shown pooled over test and retest (average test-retest correlation of 0.84) in the same style as in Fig. 3. In addition to Measured (headphone only), RAZR, and ISM, Diotic is shown on the right (yellow).

Overall externalization was rated similarly for Measured (green) and RAZR (red) for headphone representation (two upper panels). In some cases, externalization was increased for RAZR, while it was lower for pulse in the underground. Although Diotic was mostly rated "Close to the head" or "Central in the head" for headphones, as was expected, this was not always the case for the underground. Here, the high reverberation time might have made the rating difficult.

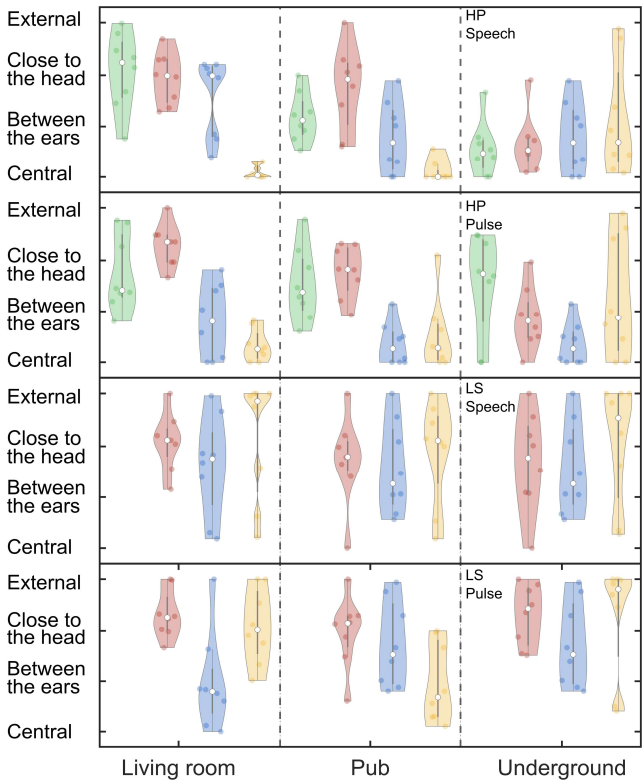

Fig. 4. Results for the externalization ratings for headphones (upper two panels; HP) ) and loudspeaker (lower two panels; LS) in the same style as in Fig. 3. Green violin plots: Measured; red plots: RAZR; blue plot: ISM; yellow plots: Diotic.

For loudspeaker presentation (lower two panels), the stimuli were somewhat more externalized, with overall similar trends as for headphones. For both, headphone and loudspeaker presentation, no systematic differences were found between speech and pulse. The only exceptions were Diotic for the living room and pub with loudspeaker. In these cases, the speech signal was judged to be more externalized than the pulse signal. ISM was perceived as less externalized, only speech with headphone playback was externalized as much as Measured and RAZR in the living room and underground.

The relatively large spread in the data indicates individual differences between participants in rating externalization. Informal interviews that were performed after the measurements, indicated that the participants found the externalization rating difficult.

Overall, a high similarity of the externalization rating was observed for Measured and RAZR using headphone presentation. Loudspeaker presentation showed slightly higher ratings on average.

## IV. Conclusion

Plausibility, overall difference, and externalization with different ALOD of the room acoustics simulation were investigated and compared in three rooms with substantially different size and (room acoustic) properties. A very high plausibility was observed for RAZR (highest ALOD applied here), comparable with the measured reference. The strongest reduction of ALOD in the shoebox ISM, was most often rated as not plausible. For speech, ISM showed higher plausibility ratings than for the pulse, suggesting that requirements for ALOD depend on the audio material. The perceived overall difference, consistent with the results of the plausibility test, was rated highest between ISM and RAZR. Despite the high similarity of RAZR and Measured in the plausibility test, however, a considerable perceived overall difference was observed between both methods. Differences between RAZR and RAZR-1st-order were minimal. For one scene and RAZR-Simple, differences between headphone and loudspeaker reproduction could be found, where elevation cues might be better represented in the loudspeaker array. Externalization was comparable for Measured and RAZR, showing no difference between speech and pulse.

Taken together, the current study shows that highly plausible simulations and similar externalization as for dummy head recordings can be achieved using strong simplifications in the acoustic simulation. Even the highest ALOD applied here (RAZR) uses a strongly simplified shoebox approximation of the room geometry for early reflections. However, effects of scattering were accounted for, and diffuse late reverberation was modelled with a physically-based FDN. Without much impact, the simulation of early reflections could be reduced from third to first order. A straight-forward implementation of a high-order shoebox ISM as often used as simplified reverberation model, performed worst, likely related to an unnatural, regular pattern of specular reflections. Further research should examine the spatial audio quality features underlying the observed overall difference ratings. Moreover, the effect on speech intelligibility is relevant for applications in hearing research.

ACKNOWLEDGMENT

The authors thank Christoph Kirsch for supporting the room acoustics simulations of this study and Bernhard U. Seeber for comments during an earlier phase of this study.